\documentclass[preprint2]{aastex}

\shorttitle{Introducing Sky in Google Earth}
\shortauthors{Scranton et al.}
\begin{document}

\title{Sky in Google Earth:\\The Next Frontier in Astronomical\\Data
Discovery and Visualization\\{\tt http://earth.google.com/sky/}}

\author{Ryan Scranton\altaffilmark{1}, Andrew Connolly\altaffilmark{2,1},
  Simon Krughoff\altaffilmark{2}, Jeremy Brewer\altaffilmark{3,1},
  Alberto Conti\altaffilmark{4}, Carol Christian\altaffilmark{4},
  Craig Sosin\altaffilmark{1}, Greg Coombe\altaffilmark{1}, 
  Paul Heckbert\altaffilmark{1}}
\email{scranton@google.com,ajc@astro.washington.edu}

\altaffiltext{1}{Google, Inc.}
\altaffiltext{2}{University of Washington, Seattle, WA 98195}
\altaffiltext{3}{University of Pittsburgh, Pittsburgh, PA 15260}
\altaffiltext{4}{Space Telescope Sciences Institute, Baltimore, MD 21218}

\begin{abstract}
Astronomy began as a visual science, first through careful
observations of the sky using either an eyepiece or the naked
eye, then on to the preservation of those images with photographic
media and finally the digital encoding of that information via
CCDs. This last step has enabled astronomy to move into a fully
automated era -- where data is recorded, analyzed and interpreted
often without any direct visual inspection. {\it Sky} in Google Earth
completes that circle by providing an intuitive visual interface to
some of the largest astronomical imaging surveys covering the full
sky. By streaming imagery, catalogs, time domain data, and ancillary
information directly to a user, {\it Sky} can provide the general public as
well as professional and amateur astronomers alike with a wealth of
information for use in education and research. We provide here a brief
introduction to {\it Sky} in Google Earth, focusing on its extensible
environment, how it may be integrated into the research process and
how it can bring astronomical research to a broader community. With an
open interface available on Linux, Mac OS X and Windows, applications
developed within {\it Sky} are accessible not just within the Google
framework but through any visual browser that supports the Keyhole
Markup Language. We present {\it Sky} as the embodiment of a virtual
telescope.
\end{abstract}


\section{Introduction}

The purpose of this paper is to provide an introduction to {\it Sky} in
Google Earth\footnote{\tt http://earth.google.com} -- describing the data
used in its creation, why certain features appear the way they do and, most
importantly, how astronomers can use {\it Sky} to create, explore, and share
their data with a broad community. We will generally avoid technical details
about the underlying mechanisms for serving the imagery and instead focus on
the operations required to place catalogs, images and educational tools within
{\it Sky} and how to serve these data sets efficiently across the web. This
paper should be viewed as a starting point for the reader's interaction with
{\it Sky} and the beginning of a discussion about what can be accomplished
with this new tool.

Astronomical data is deceptively complex. At first glance, it is
merely the result of pointing a telescope at the sky and recording
what you see.  In reality, however, a scientifically useful
description of a single observation requires knowledge of when it was
taken, from where, over what wavelengths, in what conditions, covering
what area, and so on.  Combined, the underlying images and associated
meta-data provide a data set that is not just esthetically pleasing
but is also scientifically meaningful. The goal of {\it Sky} is to create a
general framework that will enable users to access images, catalogs
and any associated meta-data across the full sky in a seamless manner
\citep{2001Sci...293.2037S}.
It will serve optical, infrared, x-rays, ultraviolet and radio images,
enable overlays of these images (with transparency), catalogs and
ancillary data associated with the underlying images, static,
time-varying and transient data. In short, it can deliver a view of the
night sky across the full electromagnetic spectrum to a user in an
efficient and scalable manner.

\section{Design}

Before discussing how to add your own data to {\it Sky}, we begin with a
description of the two principal components of {\it Sky}: the {\it basemap}
and the overlays.  The first is simply the set of images laid
out on the celestial sphere to match their positions on the sky.  The
second is the collection of lines, placemarks, and images that can be overlaid
on the base imagery and the annotations associated with the basic visual data.
Below we will describe the contents of both pieces, along
with some discussion of the design decisions that went into their
development.

\subsection{Imagery}

The core of Google Earth is a huge RGB color image pyramid projected on a
sphere, served via the internet by Google, with client software that runs on
each user's computer.  This client takes advantage of graphics hardware in
their computer to display a window into the huge image database with real time,
continuous panning and zooming.  Data is stored and served in such a way that
redisplay is as smooth as possible, giving the user the illusion that the full
detail of all the images is resident locally on their computer when, in fact,
only a small portion of it is cached there.  {\it Earth} was created for
imagery of a sphere from the outside, but for {\it Sky} we reversed that 
perspective, using the same infrastructure to serve images of space viewed
from inside the celestial sphere. 

As with Google Earth, the basemap in {\it Sky} provides a full-sphere view
of the sky by taking images with different resolutions and depths from
a range of sources, registering them to a common coordinate system and
rendering them as the user's vantage point sweeps across the sky.  For
{\it Sky} there are three primary imagery sources: the Digitized Sky Survey
(DSS; \cite{DSS}), the Sloan Digital Sky Survey (SDSS; \cite{SDSS}) and images
from the Hubble Space Telescope (HST)\footnote{\tt http://www.stsci.edu}. The
DSS is derived from plate scans of the photographic plates from the Second
Palomar Sky Survey (POSSII; \cite{POSSII}) and the UK Schmidt Southern Sky
Survey in the $J$ and $F$ passbands. The digitization of the photographic
plates was undertaken by the Space Telescope Science Institute \citep{DPOSS}
and contains a total of 1788 plates in the Northern and Southern hemispheres. 
In the north Galactic cap, the DSS images are replaced by data from the Sixth
Data Release of the Sloan Digital Sky Survey (SDSS), which covers $\sim$8000
square degrees and contains false color images generated from the $g$, $r$,
and $i$ passbands. Finally, there are 130 high resolution images drawn from
the Hubble Space Telescope which cover some of the most
interesting regions on the sky.

The union of these data sets covers the full sky. The underlying
imagery used in {\it Sky} resides in a lat/long projection \citep{HANDBOOK}.
This results in substantial distortion at the poles even when re-projecting
onto the sphere. Thus, for regions within five degrees of the pole we
replace the original images with a lower resolution view of the sky
derived from the Tycho II catalog \citep{Tycho2}. These derived images
contain stars to a depth of $B \sim 12$  and each star is represented
by a Gaussian, scaled and colored based on the star's magnitude and $B-V$
color.

\subsection{Data Registration and False Color Images}

The underlying projection and registration of images in {\it Sky} is based
on the technology used in Google Earth. This provides a mature
visualization platform on which to develop {\it Sky}, a very large user
base, a simple but extensible interface and a well-defined and on-going
support and development mechanism. It does, however, lead to a small
number of trade-offs -- related to the way geospatial data is served -- that
were made in the course of adapting the system to serve astronomical data.

The first of these concerns the bounds used for geodetic coordinates
($GEO$), which are different from those used in Equatorial coordinates
($EQ$).  The geodetic coordinates range in longitude from -180 degrees
West to +180 degrees East. Hence, there is a simple translation that
must be made from the latter to the former:
\begin{eqnarray}
RA_{\rm GEO} &=& RA_{\rm EQ} - 180^\circ \nonumber \\
DEC_{\rm GEO} &=& DEC_{\rm EQ}
\end{eqnarray}
A second issue arises due to the fact that the Earth is an oblate
spheroid which will be discussed in \S\ref{sec:image_overlays}.

All images to be served through {\it Sky} were re-projected from their
native tangent plane (or gnomic) projections onto the aforementioned
lat/long projection. SDSS images were taken from jpeg images
in the SDSS
archive\footnote{\tt http://www.sdss.org/dr6/data/das\_users\_guide\_dr2}
created as part of the standard SDSS reduction pipeline. The color of the SDSS
images was derived from the $g$, $r$ and $i$ passbands using the color
transformation proposed by \cite{LuptonRGB}. HST press release images were
taken from high resolution TIFF images created by the STScI OPO
group\footnote{\tt http://hubblesite.org}. For the DSS data, no color images
were readily available that matched the color range of the SDSS data.  Color
images were, therefore, generated from the original FITS format images
(allowing for the effects of differential chromatic abberation) in the $J$
and $F$ bands and calibrating the photographic plates such that their stellar
locus matched that of the SDSS data.  As with Google Earth, the basemap is
not static and we anticipate continued improvements as we refine our
treatment of the current imagery and new large-scale datasets become available.

\subsection{Overlays and Annotation}

\begin{figure*}
\begin{center}
\includegraphics[width=450pt]{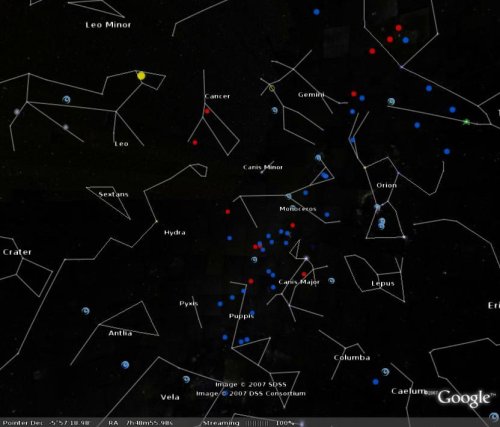}
\end{center}
\caption{\label{fig:large_scale} Large scale image of the sky showing
constellations and Messier (red) and NGC (blue) object tags.  The basemap
imagery is largely dark thanks to the re-sampling of the DSS and SDSS imagery
to lower resolution, although a number of stars are visible at the junctions
of the constellation lines.  Icons are also available for objects in the 
Yale Bright Star catalog, although they are deactivated in the current image.} 
\end{figure*}

\begin{figure*}
\begin{center}
\includegraphics[width=450pt]{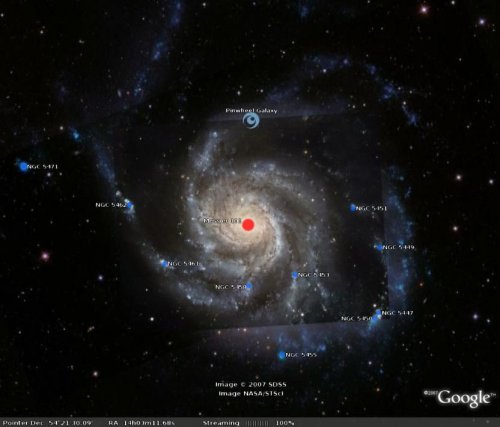}
\end{center}
\caption{\label{fig:small_scale} Closer view of the imagery for the area of
sky around M101.  As the credits at the lower middle of the image indicate,
this imagery is a combination of data from the SDSS and the HST, with the 
latter providing a higher resolution mosaic of the central galaxy region.  The
icon at the top of the M101 HST imagery contains information related to the
press release issued by the STScI for this image, as well as links to
associated papers and observational data.}
\end{figure*}

\begin{figure*}
\begin{center}
\includegraphics[width=450pt]{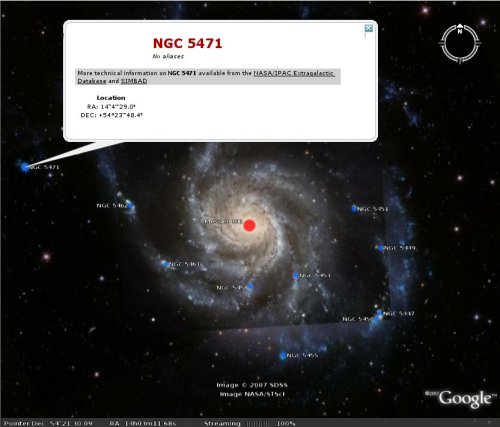}
\end{center}
\caption{\label{fig:small_scale_info} Same as Figure~\ref{fig:small_scale},
but with the information balloon for one of the NGC objects opened.  These
balloons contain basic observational data as well as links to the NED and
Simbad databases for further information and a snippet of Wikipedia text for
each object, if available.  This particular balloon, chosen so as not to
obscure the imagery, shows the bare minimum of information available for a
given object.}
\end{figure*}

As mentioned above, the built-in overlays provide a means to annotate
the basemap imagery. In its initial release {\it Sky} utilizes this
functionality to provide a simple and intuitive introduction to
objects that are visible to the general public and amateur
astronomers.  For example, we outline the 88 IAU constellations
(Figure~\ref{fig:large_scale}) provide a basic introduction to the
morphologies of galaxies visible on the sky and a description of the stages
of stellar evolution using examples in the basemap to illustrate these tours.  
Likewise, each of the images from the HST is accompanied by a pop-up balloon
containing a snippet of the press release for that image and links for further
information.  Finally, we have placed icons on objects from the Messier, New
General and Yale Bright Star catalogs (Figure~\ref{fig:small_scale}).  These
placemarks identify the object, provide basic observational data about the
source (names, positions, distance, colors and brightnesses along with links
to NED and SIMBAD where appropriate; Figure~\ref{fig:small_scale_info}) and,
in some cases, include a snippet from Wikipedia about what is known about
that star, star cluster, nebula or galaxy.

In addition to these static layers, we also provide a time-based layer
which shows the position of the Moon and Solar System planets over the
course of three months' time. The ephemeris data was generated using the
JPL Horizons interface\footnote{\tt http://ssd.jpl.nasa.gov/?horizons}.  The
associated placemarks indicate the distance, magnitude and angular extent
of each body as a function of time, controlled by a slider bar that appears
in the upper right corner of the window when these layers are activated. 
It is worth noting that planets' and the Moon's icons are not scaled according
to their actual appearance on the sky so that they can be more easily
distinguished from the background stars.  The duration of the time interval is
sufficient to show both prograde and retrograde motion in the planetary orbits.

\section{Adding Data to {\it Sky}}

While the basemap provides a common reference frame and a view of the
optical sky, the strength of {\it Sky} comes from its ability to
incorporate and display user generated data (images, catalogs, time
variable data, tours of the sky) on top of the basemap.  This wide array of 
functionality is one the advantages of a mature platform, as {\it Earth}
has been performing similar tasks for the GIS community for a number of years.
For example, the x-ray flux from a galaxy cluster can be displayed on top of
the optical galaxies seen in the basemap, which can themselves be tagged with
placemarks containing the survey data for each of the member galaxies.  
Likewise, infrared, ultra-violet and radio imaging of a single galaxy can be
overlaid simultaneously, with the user in complete control of the transparency
of each layer.  In this section, we will discuss the basic tools available for
incorporating data within {\it Sky}, as well as point to some examples
currently available for download to supplement the basic {\it Sky} package.

\subsection{Keyhole Markup Language}

Most of the methods for putting data into {\it Sky} involve some use of
Keyhole Markup Language (KML).  An introduction to the KML schema
can be found at the Google API documentation
site\footnote{\tt http://code.google.com/apis/kml/documentation/}.
Like HyperText Markup Language (HTML) and XML, KML files are composed
of tags and attributes.  Each tag describes an entity (such
as the coordinates for a point on the sky or the bounding box for an
image) and each attribute defines parameters associated with that
entity.  KML files can be used to add catalog data with placemarks,
imagery with overlays, and vector data in the form of lines and
polygons.  As with HTML, the basic building blocks can be combined in
a variety of ways to create data sets as utilitarian or richly
formatted as the user desires.  The KML standard also evolves over time,
adding features and interfaces based on user feedback and proposed usage
cases.

\subsection{Placemarks}

The most basic element for displaying data is the {\it placemark}.
This acts as a single push-pin located at a discrete point on the sky,
possibly with some associated data; indeed, the default icon for a
placemark is an image of a push-pin.  KML allows users to customize
the icon associated with a placemark, as well as the information that
appears in the pop-up balloon when a user clicks on the placemark.
More advanced features include the ability to control when the
placemark appears and how the camera moves when the user clicks on the
placemark.  For example, the appearance of the default catalog data in
{\it Sky} is tuned to the objects' magnitudes.  This keeps fainter
objects from appearing until the user is at a higher zoom level, which
avoids having the viewer overwhelmed by a unmanageable number of placemarks
on the sky at any given time.  Likewise, visibility of placemarks can be
governed by the region of the sky shown in the viewport at any given time,
preventing multiple catalogs across the full sky from being rendered at the
same time and improving the efficiency of serving the data.

\subsection{Vector Data}

The next stage of sophistication is vector data: lines and polygons.
These are essentially collections of placemarks, with line segments
connecting them and (in the latter case) a solid color filling the
enclosed area.  In both cases, users can control both the color and
transparency of the lines and polygons.  This makes vector data ideal for a
quick description of a observational footprint or uncertainty region
associated with a survey or observation, as one might have with the
observation of a gamma ray burst, for example.

\subsection{Image Overlays} \label{sec:image_overlays}

The KML tag name {\it GroundOverlay} is inherited from {\it Earth}, but
these are simply images projected against the basemap in {\it Sky}.
GroundOverlays are defined by associating an image with some bounding
region on the sky, the latter being given by the bounds in the four
cardinal directions and a rotation angle.  For small-area images, the
difference between the tangent plane and lat/long projection is slight enough
that images can be aligned to the basemap using only a linear transformation.
Google Earth has a graphical interface for aligning images this

Larger images (or those nearer the poles) can be more difficult.  As
mentioned previously, since {\it Sky} shares a rendering engine with
{\it Earth}, the geometry of the sky is, in fact, a slightly oblate spheroid 
(technically, the WGS84 projection). The GIS community has developed a
number of tools for handling re-projections of images using this geometry,
most notably the Geospatial Data Abstraction Library (GDAL).  This software
can warp images from one projection (including a tangent plane) to another
and encodes the geometric information necessary for registering the
image on the sphere in either an image header (analogous to a FITS
header) or an associated ``world file'', depending on the image
format.  While users may want to become more familiar with the GDAL
software themselves, we provide a simple, open source tool
{\it wcs2kml}\footnote{\tt http://code.google.com/p/wcs2kml/} which will read
in an image in a variety of formats and WCS information from a FITS header and
generate a properly warped image and overlay KML for you.  This tool is
available in both Python and C++ versions.

\subsection{Network Links and Regionation}\label{sec:regionation}

While the basic elements of KML are static, network links allow users to 
load new KML dynamically.  This may be triggered by a change in the view of
the camera in {\it Sky} or after a specified interval has elapsed.  This
flexibility allows for a limitless number of applications: updating the
location of satellites in orbit or their current observing target, displaying
feeds from gamma ray burst trackers, providing real-time updating of
observations to collaborators back home, and so on.  Likewise, the ability
to have network links activated based on the viewport of the sky gives
KML authors the ability to {\it regionate} their data.

{\it Regionation} is simply splitting a given data set into a hierarchical
structure, similar to what happens with the imagery and placemarks that are
part of the basic {\it Sky} data set.  When the camera is far from the 
surface of the celestial sphere, the imagery served is at a very low resolution
(finer details would be lost anyway, so displaying them is unnecessary) and
only the brightest stars and galaxies are annotated to avoid overwhelming the
user.  As the camera zooms in, higher resolution imagery is fetched
automatically and fainter objects are tagged.  The same can be accomplished
with network links to other files that are tied to progressively smaller
regions on the sky.  This allows vast amounts of catalog data and high
resolution imagery to be displayed progressively on demand, instead of
clogging the user's internet connection all at once. 
{\it Regionator}\footnote{\tt http://code.google.com/p/regionator/}, 
an open source Python program, provides a simple interface for regionating 
KML files that have already been created (like those produced by the Python
version of {\it wcs2kml}, for example).  The C++ version of {\it wcs2kml}
includes a feature which will regionate its outputs as well, producing a
hierarchical set of network linked KML files and sub-sampled versions of
astronomical images with associated WCS information.  For large mosaics (or
even images taken with large format CCDs), this can vastly improve the
performance of the resulting image overlay.  The {\it wcs2kml} package
also includes utilities for converting FITS catalog data to KML placemarks
and regionating them as well.

\subsection{Sharing \& Publishing Data}

Since KML is completely platform independent, data sharing is easy.
Simply construct a KML file describing your data and email it to
your collaborator.  Even better, place it on a webpage where it can be 
discovered and shared with the world.  Larger KML files (which are
fundamentally just simple ASCII text files) can be compressed into a
{\it zip} archive (generally tagged with a ``.kmz'' suffix to differentiate
them from uncompressed files) along with imagery to form a self-contained file
that can be opened natively by {\it Sky}.  Alternatively, imagery can be
loaded with URL pointers, just like in standard HTML.

\subsection{Working with KML in {\it Sky} mode}

KML was created to annotate Google Earth with images, placemarks and ancillary
information.  As the KML schema is designed to support this Earth-centric
annotation some of the features and tags present within KML do not translate
naturally to annotation of the sky. We provide here a brief overview of these
tags and features and provide workarounds for their use in {\it Sky}.  

To differentiate between KML for viewing on the {\it Sky} as opposed to on
{\it Earth} a hint attribute is available within the initial specification
of the KML schema. Adding the {\it Sky}
hint\footnote{\tt <kml xmlns="http://earth.google.com/kml/2.2" hint="target=sky">} to your {\tt <kml>} tag will cause the client to prompt a user to open the
KML under sky mode (if the view is currently {\it Earth}-based). All KML saved
from within the Google Earth client while it is in {\it Sky} mode will have
this attribute set by default.

Placemarks, image overlays, lines, polygons and styles are all supported under 
{\it Sky}. Tilt and roll angle for the camera position are ignored
when viewing images and placemarks under {\it Sky} as the current astronomical
imagery is a simple projection onto a sphere without altitudinal information. 
Range for the LookAt and other positional tags (i.e.~the altitude above the
Earth that the camera zooms to when a placemark is double-clicked) are
measured in meters. To convert to an angular system that is more appropriate
for viewing the sky we suggest the following relation,
\begin{equation}
R \sin(\alpha/2 - \beta/2) = (R - r) \sin(\alpha/2)
\end{equation}
where $r$ is the range to use in the LookAt tag, $R$ is the radius of the
Earth (6371 km), $\alpha$ is the angle subtended by the viewer when you are
viewing from the maximum distance from the sky (i.e.~you are pulled all the
way back to the center of the Earth; for the current client
$\alpha = 80^\circ$) and $\beta$ is the angular diameter of your region.  
For $r \ll R$, this reduces to,
\begin{equation}
\beta = 2r\tan(\alpha/2)/R
\end{equation}
where $\beta$ is in radians.

Images that are incorporated into {\it Sky} are correctly oriented (i.e.~East
to the left) and do not require any transformations other than rotation.
Rotation in {\it Sky} is in the clockwise direction (opposite from that in
{\it Earth} mode). 3D sketchup models for objects within {\it Sky} are not
officially supported but may work.

\section{Future}

The research applications of {\it Sky} are limitless.  Data can be
visualized from an all-sky view all the way to native resolution of
the underlying imagery with minimal pre-processing. Catalog data can
be immediately placed on the sky imagery with KML placemarks enabling
new discoveries to be made available in almost real time. Data can be
shared within a research group or to the broader community securely by
simply sending KML files or posting them on a webserver. Finding charts are
as simple to create as producing a screenshot or zooming to a location just
prior to observing.  Survey progress can be visualized in real time by
updating KML available over the internet (via network links) as images come
off the CCDs. Finally, when the observations are ready to be made public,
{\it Sky} is an ideal platform to reach parts of the general public interested
in astronomy and put your results in the context of the broader astrophysical
data universe.

As a tool with a simple, extensible and intuitive interface we see {\it Sky}
in Google Earth as the embodiment of a Virtual Telescope, providing
open access to data through open standards. It leverages the power of
the Google infrastructure for serving data to a wide range of people
while also enabling any user to integrate and share their own images
and catalogs. We hope that this step towards the democratization of
science will enable a new era of visual discovery and science
communication.

\acknowledgments

The authors wish to thank Chikai Ohazama, Lior Ron, Effie Seiberg, Steve
Zelinka, Andrew Moore, Robert Pike, Brian McClendon, Linne Ha, Mohammad Khan,
Brian McClean, Dylan Myers, Wei Luo, Peter Birch, Andrew Kirmse, Michael
Ashbridge, Bent Hagemark, Chris DiBona, Sam Roweis, and Keir Mierle for their
tireless efforts to make {\it Sky} what it is and what it will be.

Please see the {\it Sky} partners
homepage\footnote{\tt http://earth.google.com/sky/partners.html} for a list
of the institutions contributing imagery to {\it Sky}.

\end{document}